\documentclass[10pt,prd,twocolumn,showpacs,groupedaddress,superscriptaddress,amsmath,amssymb,amsfonts]{revtex4-2}

\usepackage[T1]{fontenc}

\usepackage{lineno,hyperref}
\usepackage{upgreek}
\usepackage{xcolor}
\usepackage{lipsum}
\usepackage{orcidlink}
\usepackage{siunitx}
\usepackage{soul}
\usepackage{bigdelim,bigdelim, makecell, booktabs}
\modulolinenumbers[1]

\usepackage{graphicx}

\usepackage{comment}

\newcommand{\Apr}{A^\prime}
\newcommand{\Zpr}{Z^\prime}
\newcommand{\mZ}{m_{\Zpr}}
\newcommand{\gZ}{g_{\Zpr}}
\newcommand\UBL{U(1)_{B-L}}
\newcommand\alphaD{\alpha_{D}}
\newcommand\gBL{g_{B-L}}
\newcommand{\Sphys}{\mathcal{S}_{phys}}

\definecolor{colorOrange}{HTML}{ffa070}
\definecolor{colorRed}{HTML}{c1244f}
\definecolor{colorLightRed}{HTML}{ff5781}
\definecolor{colorDarkRed}{HTML}{94003a}
\definecolor{colorRed}{HTML}{c1244f}
\definecolor{colorLightBlue}{HTML}{add8e6}
\definecolor{colorYellow}{HTML}{ffa500}
\definecolor{colorPink}{HTML}{ff00ff}
\definecolor{colorGreen}{HTML}{33cd32}

\begin{document}

\title{Improved limits on a new \texorpdfstring{\ensuremath{\Zpr}}{Zpr} in \texorpdfstring{\ensuremath{B-L}}{B-L} scenarios with the NA64 experiment at CERN}
\thanks{\textcopyright 2026 CERN for the benefit of the NA64 Collaboration.
Reproduction of this article or parts of it is allowed as specified in the CC-BY-4.0 license.}
\author{Yu.~M.~Andreev\orcidlink{0000-0002-7397-9665}}
\affiliation{Authors affiliated with an institute covered by a cooperation agreement with CERN}
\author{A.~Antonov\orcidlink{0000-0003-1238-5158}}
\affiliation{INFN, Sezione di Genova, 16147 Genova, Italia}
\author{M.~A.~Ayala~Torres\orcidlink{0000-0002-4296-9464}}
\affiliation{Center for Theoretical and Experimental Particle Physics, Facultad de Ciencias Exactas, Universidad Andres Bello, Fernandez Concha 700, Santiago, Chile}
\affiliation{Millennium Institute for Subatomic Physics at High-Energy Frontier (SAPHIR), Fernandez Concha 700, Santiago, Chile}
\author{D.~Banerjee\orcidlink{0000-0003-0531-1679}}
\affiliation{CERN, European Organization for Nuclear Research, CH-1211 Geneva, Switzerland}
\author{B.~Banto Oberhauser\orcidlink{0009-0006-4795-1008}}
\email[\textbf{Corresponding authors:}]{bantoobb@ethz.ch,}
\affiliation{ETH Z\"urich, Institute for Particle Physics and Astrophysics, CH-8093 Z\"urich, Switzerland}
\author{V.~Bautin\orcidlink{0000-0002-5283-6059}}
\affiliation{Authors affiliated with an international laboratory covered by a cooperation agreement with CERN}
\author{J.~Bernhard\orcidlink{0000-0001-9256-971X}}
\affiliation{CERN, European Organization for Nuclear Research, CH-1211 Geneva, Switzerland}
\author{P.~Bisio\orcidlink{/0009-0006-8677-7495}}
\affiliation{Universit\`a degli Studi di Genova, 16126 Genova, Italia}
\affiliation{INFN, Sezione di Genova, 16147 Genova, Italia}
\author{A.~Celentano\orcidlink{0000-0002-7104-2983}}
\affiliation{INFN, Sezione di Genova, 16147 Genova, Italia}
\author{N.~Charitonidis\orcidlink{0000-0001-9506-1022}}
\affiliation{CERN, European Organization for Nuclear Research, CH-1211 Geneva, Switzerland}
\author{P.~Crivelli\orcidlink{0000-0001-5430-9394}}
\email{crivelli@phys.ethz.ch}
\affiliation{ETH Z\"urich, Institute for Particle Physics and Astrophysics, CH-8093 Z\"urich, Switzerland}
\author{A.~V.~Dermenev\orcidlink{0000-0001-5619-376X}}
\affiliation{Authors affiliated with an institute covered by a cooperation agreement with CERN}
\author{S.~V.~Donskov\orcidlink{0000-0002-3988-7687}}
\affiliation{Authors affiliated with an institute covered by a cooperation agreement with CERN}
\author{R.~R.~Dusaev\orcidlink{0000-0002-6147-8038}}
\affiliation{Authors affiliated with an institute covered by a cooperation agreement with CERN}
\author{T.~Enik\orcidlink{0000-0002-2761-9730}}
\affiliation{Authors affiliated with an international laboratory covered by a cooperation agreement with CERN}
\author{V.~N.~Frolov}
\affiliation{Authors affiliated with an international laboratory covered by a cooperation agreement with CERN}
\author{S.~V.~Gertsenberger\orcidlink{0009-0006-1640-9443}}
\affiliation{Authors affiliated with an international laboratory covered by a cooperation agreement with CERN}
\author{S.~Girod}
\affiliation{CERN, European Organization for Nuclear Research, CH-1211 Geneva, Switzerland}
\author{S.~N.~Gninenko\orcidlink{0000-0001-6495-7619}}
\affiliation{Authors affiliated with an institute covered by a cooperation agreement with CERN}
\author{M.~H\"osgen}
\affiliation{Universit\"at Bonn, Helmholtz-Institut f\"ur Strahlen-und Kernphysik, 53115 Bonn, Germany}
\author{Y.~Kambar\orcidlink{0009-0000-9185-2353}}
\affiliation{Authors affiliated with an international laboratory covered by a cooperation agreement with CERN}
\author{A.~E.~Karneyeu\orcidlink{0000-0001-9983-1004}}
\affiliation{Authors affiliated with an institute covered by a cooperation agreement with CERN}
\author{G.~Kekelidze\orcidlink{0000-0002-5393-9199}}
\affiliation{Authors affiliated with an international laboratory covered by a cooperation agreement with CERN}
\author{B.~Ketzer\orcidlink{0000-0002-3493-3891}}
\affiliation{Universit\"at Bonn, Helmholtz-Institut f\"ur Strahlen-und Kernphysik, 53115 Bonn, Germany}
\author{D.~V.~Kirpichnikov\orcidlink{0000-0002-7177-077X}}
\affiliation{Authors affiliated with an institute covered by a cooperation agreement with CERN}
\author{M.~M.~Kirsanov\orcidlink{0000-0002-8879-6538}}
\affiliation{Authors affiliated with an institute covered by a cooperation agreement with CERN}
\author{V.~A.~Kramarenko\orcidlink{0000-0002-8625-5586}}
\affiliation{Authors affiliated with an international laboratory covered by a cooperation agreement with CERN}
\affiliation{Authors affiliated with an institute covered by a cooperation agreement with CERN}
\author{L.~V.~Kravchuk\orcidlink{0000-0001-8631-4200}}
\affiliation{Authors affiliated with an institute covered by a cooperation agreement with CERN}
\author{N.~V.~Krasnikov\orcidlink{0000-0002-8717-6492}}
\affiliation{Authors affiliated with an international laboratory covered by a cooperation agreement with CERN}
\affiliation{Authors affiliated with an institute covered by a cooperation agreement with CERN}
\author{S.~V.~Kuleshov\orcidlink{0000-0002-3065-326X}}
\affiliation{Millennium Institute for Subatomic Physics at High-Energy Frontier (SAPHIR), Fernandez Concha 700, Santiago, Chile}
\affiliation{Center for Theoretical and Experimental Particle Physics, Facultad de Ciencias Exactas, Universidad Andres Bello, Fernandez Concha 700, Santiago, Chile}
\author{V.~E.~Lyubovitskij\orcidlink{0000-0001-7467-572X}}
\affiliation{Millennium Institute for Subatomic Physics at High-Energy Frontier (SAPHIR), Fernandez Concha 700, Santiago, Chile}
\author{V.~Lysan\orcidlink{0009-0004-1795-1651}}
\affiliation{Authors affiliated with an international laboratory covered by a cooperation agreement with CERN}
\author{A.~Marini\orcidlink{0000-0002-6778-2161}}
\affiliation{INFN, Sezione di Genova, 16147 Genova, Italia}
\author{L.~Marsicano\orcidlink{0000-0002-8931-7498}}
\affiliation{INFN, Sezione di Genova, 16147 Genova, Italia}
\author{V.~A.~Matveev\orcidlink{0000-0002-2745-5908}}
\affiliation{Authors affiliated with an international laboratory covered by a cooperation agreement with CERN}
\author{R.~Mena~Fredes}
\affiliation{Millennium Institute for Subatomic Physics at High-Energy Frontier (SAPHIR), Fernandez Concha 700, Santiago, Chile}
\author{R.~Mena~Yanssen}
\affiliation{Millennium Institute for Subatomic Physics at High-Energy Frontier (SAPHIR), Fernandez Concha 700, Santiago, Chile}
\affiliation{Universidad T\'ecnica Federico Santa Mar\'ia and CCTVal, 2390123 Valpara\'iso, Chile}
\author{L.~Molina Bueno\orcidlink{0000-0001-9720-9764}}
\affiliation{Instituto de Fisica Corpuscular (CSIC/UV), Carrer del Catedratic Jose Beltran Martinez, 2, 46980 Paterna, Valencia, Spain}
\author{M.~Mongillo\orcidlink{0009-0000-7331-4076}}
\affiliation{ETH Z\"urich, Institute for Particle Physics and Astrophysics, CH-8093 Z\"urich, Switzerland}
\author{D.~V.~Peshekhonov\orcidlink{0009-0008-9018-5884}}
\affiliation{Authors affiliated with an international laboratory covered by a cooperation agreement with CERN}
\author{V.~A.~Polyakov\orcidlink{0000-0001-5989-0990}}
\affiliation{Authors affiliated with an institute covered by a cooperation agreement with CERN}
\author{B.~Radics\orcidlink{0000-0002-8978-1725}}
\affiliation{York University, Toronto, Canada}
\author{K.~Salamatin\orcidlink{0000-0001-6287-8685}}
\affiliation{Authors affiliated with an international laboratory covered by a cooperation agreement with CERN}
\author{V.~D.~Samoylenko}
\affiliation{Authors affiliated with an institute covered by a cooperation agreement with CERN}
\author{H.~Sieber\orcidlink{0000-0003-1476-4258}}
\affiliation{ETH Z\"urich, Institute for Particle Physics and Astrophysics, CH-8093 Z\"urich, Switzerland}
\author{D.~Shchukin\orcidlink{0009-0007-5508-3615}}
\affiliation{Authors affiliated with an institute covered by a cooperation agreement with CERN}
\author{O.~Soto}
\affiliation{Millennium Institute for Subatomic Physics at High-Energy Frontier (SAPHIR), Fernandez Concha 700, Santiago, Chile}
\affiliation{Departamento de Fisica, Facultad de Ciencias, Universidad de La Serena, Avenida Cisternas 1200, La Serena, Chile}
\author{V.~O.~Tikhomirov\orcidlink{0000-0002-9634-0581}}
\affiliation{Authors affiliated with an institute covered by a cooperation agreement with CERN}
\author{I.~Tlisova\orcidlink{0000-0003-1552-2015}}
\affiliation{Authors affiliated with an institute covered by a cooperation agreement with CERN}
\author{A.~N.~Toropin\orcidlink{0000-0002-2106-4041}}
\affiliation{Authors affiliated with an institute covered by a cooperation agreement with CERN}
\author{M.~Tuzi\orcidlink{0009-0000-6276-1401}}
\affiliation{Instituto de Fisica Corpuscular (CSIC/UV), Carrer del Catedratic Jose Beltran Martinez, 2, 46980 Paterna, Valencia, Spain}
\author{P.~V.~Volkov\orcidlink{0000-0002-7668-3691}}
\affiliation{Authors affiliated with an international laboratory covered by a cooperation agreement with CERN}
\affiliation{Authors affiliated with an institute covered by a cooperation agreement with CERN}
\author{I.~V.~Voronchikhin\orcidlink{0000-0003-3037-636X}}
\affiliation{Authors affiliated with an institute covered by a cooperation agreement with CERN}
\author{J.~Zamora-Sa\'a\orcidlink{0000-0002-5030-7516}}
\affiliation{Millennium Institute for Subatomic Physics at High-Energy Frontier (SAPHIR), Fernandez Concha 700, Santiago, Chile}
\affiliation{Center for Theoretical and Experimental Particle Physics, Facultad de Ciencias Exactas, Universidad Andres Bello, Fernandez Concha 700, Santiago, Chile}
\author{A.~S.~Zhevlakov\orcidlink{0000-0002-7775-5917}}
\affiliation{Authors affiliated with an international laboratory covered by a cooperation agreement with CERN}

\begin{abstract}
Extensions of the Standard Model featuring an additional $U(1)_{B-L}$ gauge symmetry provide a compelling framework linking the origin of neutrino masses to possible dark matter candidates. The associated gauge boson, $\Zpr$, couples directly to Standard Model fermions and can be produced in fixed-target experiments through electron-nucleus interactions. 
In this work, we present new constraints on the coupling constant $g_{B-L}$ obtained with the NA64 experiment using the full electron-beam dataset collected between 2016 and 2022, corresponding to $(9.4\pm0.5)\times10^{11}$ electrons on target. 
The analysis includes the resonant $e^{+}e^{-}$ annihilation production channel, which enhances sensitivity in the mass range $\mZ\in[200,300]~\mathrm{MeV}$. The larger dataset provides approximately three times the statistics of previous analyses, thereby improving sensitivity. For the unbroken $U(1)_{B-L}$ case, the new limits exceed those from dedicated neutrino-scattering experiments, providing the most stringent laboratory bounds on $g_{B-L}$ for sub-GeV masses of the new boson. In scenarios where the $\Zpr$ couples to dark matter, the decay width is dominated by invisible channels, and the corresponding exclusion limits can be directly derived from the NA64 invisible-mode analysis.
\end{abstract}

\maketitle

\section{Introduction}\label{sec:Introduction}

A wide class of Standard Model (SM) extensions features an additional $U(1)'$ gauge symmetry, offering well-motivated frameworks for addressing open questions such as the origin of neutrino masses and the nature of dark matter (DM). Among these, models based on the baryon minus lepton number $B-L$ symmetry stand out for their simplicity and predictive power. Such models have recently received significant attention, as they can naturally accommodate right-handed neutrinos (RHNs) and provide viable DM candidates~\cite{Langacker:2008yv, Heeck:2014zfa, Okada:2018ktp, Ilten:2018crw, Okada:2020cue, NA64:2022yly, Asai:2022zxw, Asai:2023mzl, KA:2023dyz, A:2025ygb, Komachenko:1989qn, Fayet:1980ad, Fayet:1980rr, Fayet:1990wx, Fayet:2004bw, Boehm:2003hm, Fayet:2007ua}. 

In $\UBL$ models, the inclusion of RHNs is required to cancel gauge anomalies, which in turn allows interactions between the SM and these otherwise sterile states. In contrast, anomaly cancellation is achieved without introducing RHNs in models gauging lepton flavor differences, such as $L_{\mu}-L_{\tau}$~\cite{He:1990pn, He:1991qd}. Anomaly cancellation in $B-L$ models can occur in two distinct ways, depending on whether neutrinos are Majorana or Dirac particles. In the Majorana case, the symmetry is explicitly broken by two units, $\Delta(B-L)=2$, allowing for baryon and lepton number violating processes such as neutrinoless double beta decay~\cite{FileviezPerez:2022ypk, LEGEND:2025jwu}. In contrast, if neutrinos are Dirac particles, the $\UBL$ symmetry remains unbroken, requiring three light RHNs to ensure anomaly cancellation.

These extensions predict a new massive vector boson, denoted as $\Zpr$. Unlike the dark photon, a $\UBL$ gauge boson couples directly to SM fermions (neutrinos $\nu_{l}$, charged leptons $l$ with $l = e, \mu,\tau$, and quarks $q=u,d$), as all carry $B-L$ charge. The interaction Lagrangian in the unbroken $\UBL$ case can be written as
\begin{multline}
\mathcal{L} \supset \gZ\, Z'_{\mu} 
\Big[
\sum_{l=e,\mu,\tau} Q'_{l}\, \left(\bar{\nu}_{l}\gamma^{\mu}\nu_{l}
+ \bar{l}\gamma^{\mu}l\right) \\
+ \sum_{q=u,d} Q'_{q}\, \bar{q}\gamma^{\mu}q
\Big],
\label{eq:theory-BminusL-interaction-lagrangian}
\end{multline}
where $\gZ=\gBL=\sqrt{4\pi\alpha_{B-L}}$ is the coupling and $Q_{i}^{\prime}$ are the $U(1)'$ charges of SM fermions. Several anomaly-free combinations of baryon and lepton numbers, denoted as $B-n_{i}L_{i}$, lead to different couplings to charged leptons. Table~\ref{tab:theory-BminusL-U1prime_models} summarizes the models most relevant to electron-beam searches.

\begin{table}[ht]
    \centering
    \begin{tabular}{|c|cc|ccc|}
    \hline
    Model & $Q_{u}^{\prime}$ & $Q_{d}^{\prime}$ & $Q_{e}^{\prime}$ & $Q_{\mu}^{\prime}$ & $Q_{\tau}^{\prime}$ \\
    \hline\hline
    $B-L$ & $1/3$ & $1/3$ & $-1$ & $-1$ & $-1$ \\ 
    $B-3L_{e}$ & $1/3$ & $1/3$ & $-3$ & $0$ & $0$ \\ 
    $B-2L_{e}-L_{\mu}$ & $1/3$ & $1/3$ & $-2$ & $-1$ & $0$ \\ 
    $B-L_{e}-2L_{\mu}$ & $1/3$ & $1/3$ & $-1$ & $-2$ & $0$ \\
    \hline
    \end{tabular}
\caption{Anomaly-free $U(1)'$ models and corresponding fermion charges relevant for electron-beam experiments.}
\label{tab:theory-BminusL-U1prime_models}
\end{table}

Since $\UBL$ models predict couplings to both charged leptons and neutrinos, they can be probed through precision measurements of the neutrino-electron scattering cross section. Consequently, strong constraints arise from solar~\cite{XMASS:2020zke, CDEX:2022mlp}, accelerator~\cite{AtzoriCorona:2022moj}, and reactor~\cite{Lindner:2018kjo, Bilmis:2015lja} neutrino experiments, where new neutral-current interactions could enhance scattering rates beyond SM predictions.

\begin{figure}[ht!]
\centering
\includegraphics[width=0.49\columnwidth]{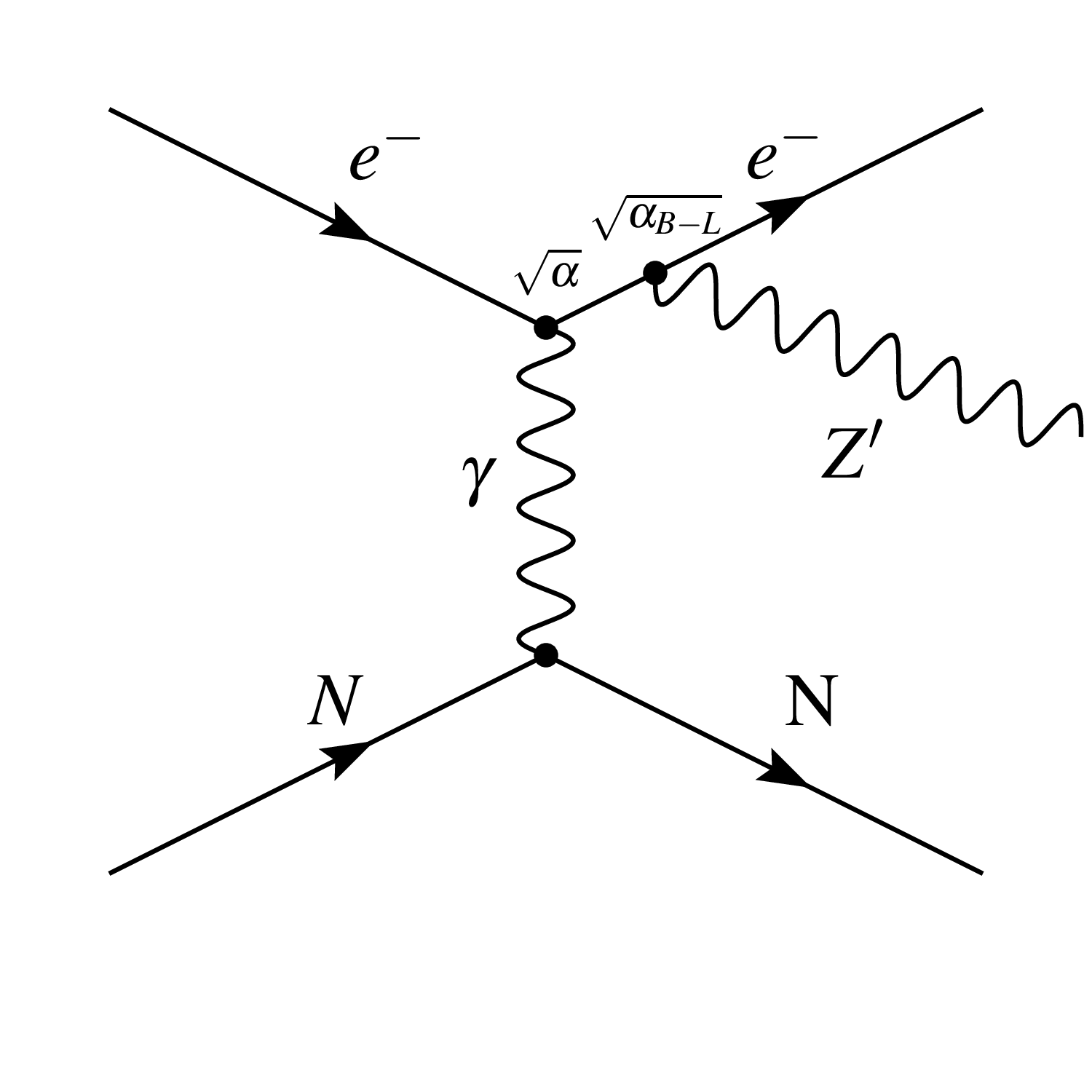}\hfill
\includegraphics[width=0.49\columnwidth]{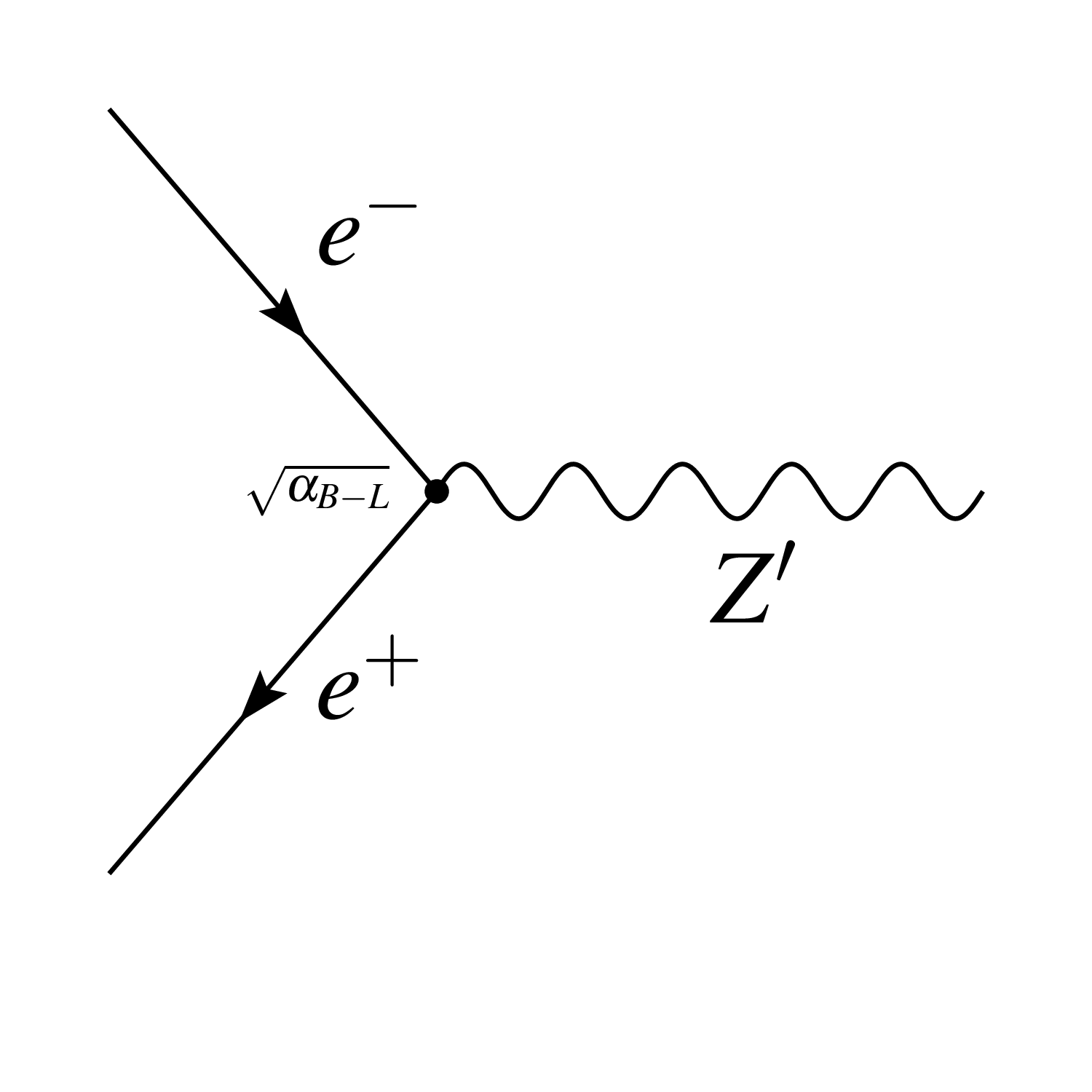}
\caption{Main $\Zpr$ production mechanisms in fixed-target experiments.}
\label{fig:theory-BminusL-Zpr_feynman}
\end{figure}

In fixed-target experiments such as NA64, the $\Zpr$ can be produced in electron-nucleus interactions via its tree-level coupling to SM fermions. The dominant production mechanisms, analogous to the dark photon case, are Bremsstrahlung and resonant $e^{+}e^{-}$ annihilation, as illustrated in Fig.~\ref{fig:theory-BminusL-Zpr_feynman}. The cross sections of these processes, along with the $\Zpr$ decay rates, can be derived from the dark photon case by replacing the kinetic mixing parameter $\varepsilon e$ with the $\UBL$ coupling $\gBL$~\cite{Fabbrichesi:2020wbt}.

\begin{figure*}[ht!]
\begin{center}
\includegraphics[trim={0 0 1cm 0}, clip, width=0.99\textwidth]{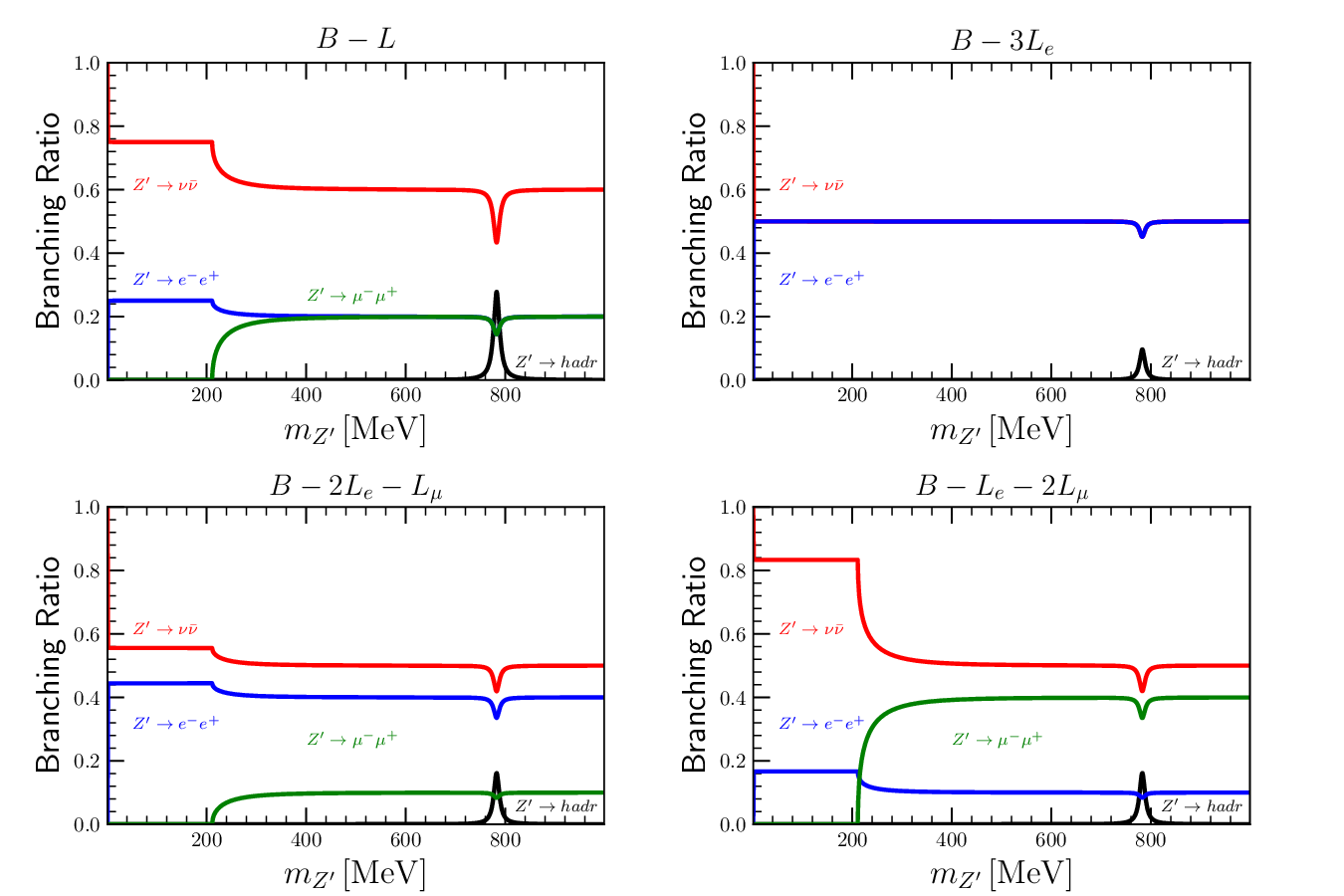}
\caption{Branching fractions of $\Zpr$ decay channels for $\mZ < 1~\mathrm{GeV}$ in the unbroken symmetry case (Dirac neutrinos) described by Eq.~\ref{eq:theory-BminusL-interaction-lagrangian} for various $B-L$-type models.}
\label{fig:theory-BminusL-Zpr_decay_BR} 
\end{center}
\end{figure*}

\begin{figure}[ht!]
\begin{center}
\includegraphics[trim={0.5cm 0 0.5cm 0}, clip, width=0.99\columnwidth]{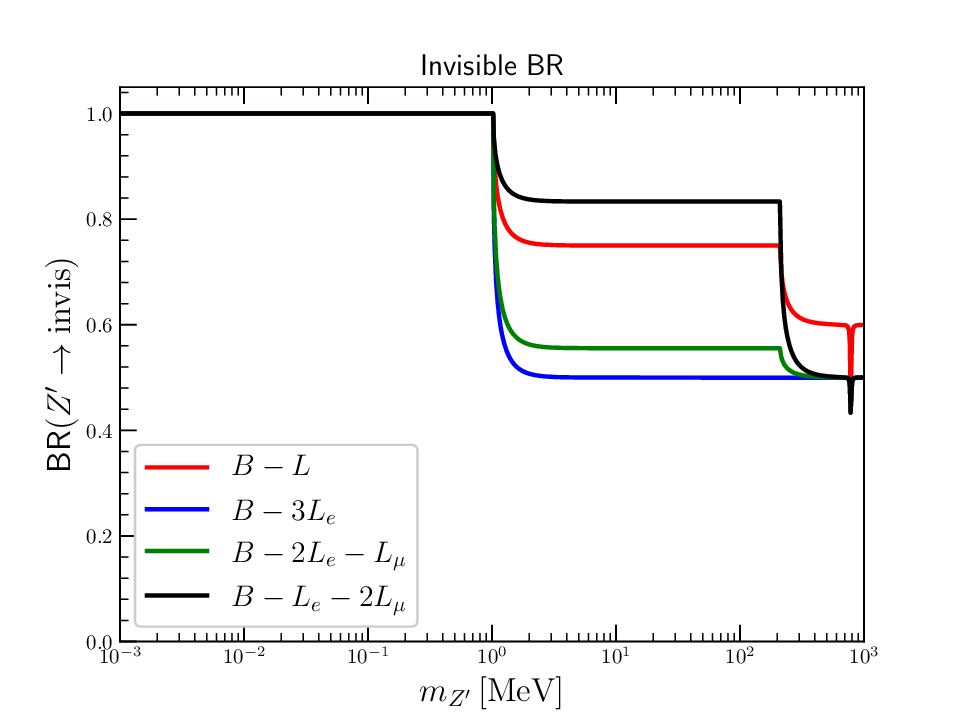}
\caption{Invisible branching ratio as a function of mediator mass for different $B-L$-type models in the unbroken symmetry (Dirac neutrinos).}
\label{fig:theory-BminusL-Zpr_invis_BR} 
\end{center}
\end{figure}

The $\Zpr$ total decay width \footnote{See Appendix~\ref{sec:appendix-Zpr-resonant} for expressions of the partial widths.} is obtained by summing over all possible final states. For $\mZ < 1~\mathrm{GeV}$, decays to both leptonic and hadronic channels are kinematically allowed, with the relative fractions depending on the specific charge assignment. Below the $e^{+}e^{-}$ threshold, $\mZ < 2m_{e}$, the $\Zpr$ decays predominantly to neutrinos, yielding an invisible signature. As the mediator mass increases, visible final states become accessible. However, hadronic decay modes are subdominant over most of the mass range below the $\omega$ meson mass ($m_{\omega}\approx 783~\mathrm{MeV}$)~\cite{Heeck:2014zfa}. This is because the $\Zpr$ is an isoscalar, making the primary hadronic decay $\Zpr \rightarrow \pi\pi$ isospin violating and therefore highly suppressed. The corresponding branching fractions and invisible ratios are shown in Figs.~\ref{fig:theory-BminusL-Zpr_decay_BR} and~\ref{fig:theory-BminusL-Zpr_invis_BR} for the unbroken symmetry scenario.

A particularly compelling aspect of $B-L$ models is their sensitivity to the underlying neutrino mass-generation mechanism. In the broken $\UBL$ case, neutrinos acquire Majorana masses via the type-I seesaw mechanism, introducing heavy right-handed states that modify the invisible decay width of the $\Zpr$. These states reduce the invisible branching fraction and enhance visible decay channels when compared to the unbroken scenario presented in Fig.~\ref{fig:theory-BminusL-Zpr_invis_BR}, making the ratio of visible to invisible modes a potential probe of the neutrino sector~\cite{Heeck:2014zfa}. Furthermore, if the $\Zpr$ decay width can be determined experimentally from resonant annihilation, it could offer a novel window into neutrino mass generation.

Additionally, introducing a dark sector coupling provides strong motivation for the $B-L$ $\Zpr$ search. Assuming the dark matter charge heavily dominates over the Standard Model charges ($Q_{\chi}^{\prime}\gg Q_{i}^{\prime}$), the $\Zpr$ would in this case decay predominantly invisibly into a pair of light Dark Matter (LDM) particles. This results in a missing-energy signature akin to that of an invisible dark photon, while naturally supporting LDM as a thermal relic with distinct parameter-space targets.

Experimental sensitivities to the $\Zpr$ can be reinterpreted across the models listed in Table~\ref{tab:theory-BminusL-U1prime_models} by rescaling the fermion charges $Q_{i}^{\prime}$. The radiative production rate scales approximately as $\sigma^{\mathrm{rad}}\propto Q_{e}^{\prime 2}\times \mathrm{Br}(\Zpr\rightarrow \mathrm{invis})$, as indicated by the vertex factors in Fig.~\ref{fig:theory-BminusL-Zpr_feynman}. Consequently, relative to the canonical $B-L$ model, the expected invisible signal yield from radiative production is enhanced by a factor of nine for a $B-3L_{e}$ gauge boson when $\mZ < 2m_{e}$, and by a factor of $6.0$ for $2m_{e} \ll \mZ$, once the decay $\Zpr\rightarrow e^+e^-$ becomes kinematically allowed. In contrast, the expected number of invisible signal events from resonant annihilation scales with the partial decay width to electrons $\sigma^{\mathrm{annih}}\propto \Gamma_{e^+e^-}\times \mathrm{Br}(\Zpr\rightarrow \mathrm{invis})$, as discussed in Appendix~\ref{sec:appendix-Zpr-resonant}.

In this work, we present new constraints on the coupling strength $g_{B-L}$ obtained from the NA64 experiment, probing the sub-GeV mass range of the benchmark scenario of a new $\UBL$ gauge boson. These results extend existing limits and provide complementary sensitivity to other laboratory and astrophysical searches, thereby narrowing the parameter space for $B-L$ extensions of the SM.

\section{\texorpdfstring{\ensuremath{\Zpr}}{Zpr} searches at NA64}\label{sec:na64}

NA64 is a fixed-target experiment exploiting the high-energy beams in the North Area facility at CERN. In particular, a high-energy electron beam is produced by impinging $\SI{400}{GeV/c}$ protons on a thick beryllium target. Secondary photons, originating predominantly from $\pi^0 \rightarrow \gamma\gamma$ decays of neutral pions generated in the target, are subsequently converted into $e^+e^-$ pairs in a thin Pb foil \cite{Atherton:1980vj, Atherton:164934}. The resulting tertiary electrons and positrons are then selected by a septum dipole magnet based on their momentum and charge. During nominal operation, the SPS delivers a single spill of $\mathcal{O}(10^{13})$ protons to the North Area every $\SI{10.8}{s}$ cycle, with each spill lasting $\SI{4.8}{s}$ \cite{Prebibaj:2848908}. This results in $\mathcal{O}(10^{7})$ electrons per spill, which are then transported through a $\SI{540}{m}$-long evacuated beamline to the experimental area in the EHN1 hall \cite{Banerjee:2774716}. During the 2022 data-taking period, the experiment operated at a beam rate up to $\SI{2}{\mega \hertz}$. For this period, the intrinsic hadron contamination, mainly consisting of kaons and pions, was estimated to be below the percent level ($h/e \lesssim 1\%$) \cite{Andreev:2023xmj}.

The electron beam is irradiated onto an active beam dump that serves as the target where the production of a new mediator, such as the $\Zpr$, takes place. This particle can then promptly decay into neutrinos or a pair of LDM particles, carrying away some of the initial electron's energy and leaving the setup undetected, owing to their feeble interactions with SM matter. Therefore, the signature in the invisible searches at NA64 is characterized by missing energy from a clean electron primary impinging on the target. The presence of these new particles is inferred from the energy deficit they exhibit.

An alternative approach adopted in traditional beam-dump experiments \cite{deNiverville:2011it} is to observe the scattering of the resulting DM particles in a downstream detector, akin to neutrino neutral-current scattering. Since this method relies on an additional interaction of the DM particles in the detector, it reduces the overall sensitivity of these experiments. In contrast, a key advantage of NA64 is that the signal yield scales only with the production cross section, which is proportional to the square of the coupling between SM particles and the new force mediator. In the case of a massive $\UBL$ gauge boson, this can be formulated as:
\begin{equation}
    \underbrace{\sigma \propto \gBL^{2}}_{\mathrm{Missing\ energy\ }\mathbf{(NA64)}}\quad \mathrm{vs.}\quad\quad
    \underbrace{\sigma \propto \gBL^{4}\times \alphaD}_{\mathrm{DM\ scattering}} 
    \label{eq:na64-sensitivity-comparison}
\end{equation}

For the small coupling motivated by thermal LDM, $\gBL\approx10^{-4}$, NA64 requires roughly two orders of magnitude fewer statistics than traditional beam-dump experiments to reach a comparable sensitivity.

\section{Experimental setup}\label{sec:ExpSetup}

\begin{figure*}[t!]
    \centering
    \includegraphics[width=.99\textwidth]{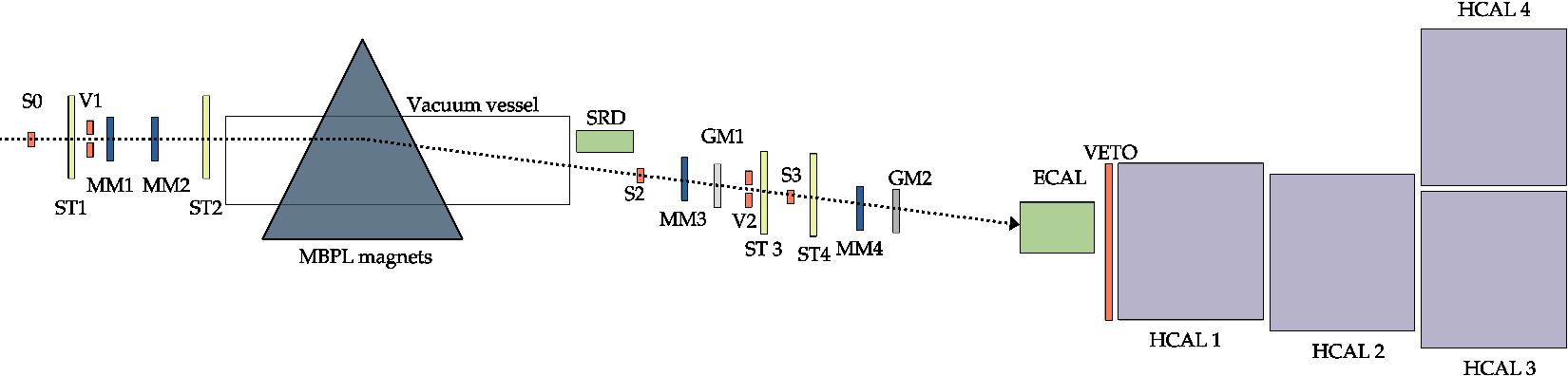}
    \caption{The NA64 setup in the invisible mode configuration during the 2022 data taking at the H4 beamline. See text for further details.}
    \label{fig:na64-subdetectors-2023-setup}
\end{figure*}

Searching for rare events at NA64 relies on properly defining the incoming electron and detecting the secondary particles produced in the target or the material along the beam. In addition to the active thick-target, several detectors are in place to control the background expected to mimic the invisible signature, minimizing the possibility of a false positive. For this purpose, the NA64 detector, as shown in Fig.~\ref{fig:na64-subdetectors-2023-setup} consists of the following sub-systems:

\begin{itemize}
  \item[(I)] A series of scintillator counters (Sc) that define the beam and trigger conditions for the data acquisition. This includes three $\SI{1.5}{mm}$ thin, $\SI{32}{mm}\varnothing$ circular scintillators $S_{0}$, $S_{2}$, $S_{3}$, and two $\SI{10}{mm}$ thick scintillator plates with $\SI{40}{mm}\varnothing$ holes, $V_{1}$, $V_{2}$.
    \item[(II)] A magnetic spectrometer to reconstruct the incoming momenta of the $\SI{100}{GeV}$ beam electrons, with Straw chambers \cite{Volkov:2019qhb} (ST), Gas Electron Multiplier (GEM) and Micromegas (MM) detectors \cite{Banerjee:2017mdu}. Two consecutive large-gap uniform-field dipole magnets (MBPL) deflect the charged beam particles, and a $\SI{16}{m}$-long vacuum tube connects the upstream and downstream sections of the setup. Each MBPL magnet is $\SI{2}{m}$ long, and together they provide an integrated field equivalent to $\simeq\SI{7}{T m}$. The resulting deflection is approximately $\SI{22}{mrad}$ and the momentum resolution is of the order $\delta p /p_{reco}\approx\mathcal{O}(1\%)$ \cite{Banerjee:2017mdu}.
    \item[(III)] A synchrotron radiation detector (SRD) to tag incoming electrons based on their emission of SR photons as their trajectory is bent by the magnetic field \cite{Depero:2017mrr}. The detector consists of three arrays of alternating Pb and Sc layers, positioned between the undeflected and deflected beams. Given that the SR energy scales as $E^4_{0}/m^{4}$ for a charged particle with mass $m$ and energy $E_{0}$ \cite{Woodruff_2021}, the SRD is an effective tool to reject low-energy electrons and heavier beam contaminants such as muons or hadrons.
    \item[(IV)] A $19\times23\times\SI{47}{cm^3}$, $40$ radiation lengths ($X_{0}$) Pb/Sc electromagnetic calorimeter (ECAL) that works as the active target. The first $4X_{0}$ are read out separately and serve as the pre-shower (PRS) detector. The ECAL is transversely segmented into a $5\times6$ matrix of cells to measure the transverse shower shape, which is crucial for discriminating against hadronic showers. Each cell consists of 150 layers of $\SI{1.5}{mm}$ Pb and $\SI{1.5}{mm}$ Sc material. The ECAL's energy resolution is approximately $\sigma_{E}/E \simeq 10\%/\sqrt{E(\mathrm{GeV})}\oplus 4\%$ \cite{Andreev:2023xmj}.
    \item[(V)] A large, high-efficiency plastic scintillator counter (VETO) that rejects charged particles leaking out of the ECAL. It is subdivided into three vertical slabs, each read by a pair of PMTs.
    \item[(VI)] Four $60\times60\times\SI{163}{cm^3}$ hadronic calorimeters (HCAL) that, given the large Lorentz boost of the incoming electrons, provide the necessary hermeticity to ensure the detection of hadronic or muon contaminants as well as secondary hadrons produced in the target. Three modules are placed along the deflected beamline, while HCAL4 is placed at zero degrees to catch the fraction of neutral particles in the beam \footnote{In the following, the HCAL4 module is treated independently from the other modules. Therefore, unless stated otherwise, any cuts on the \emph{total} HCAL energy deposition refer exclusively to the first three modules. }. Each module consists of Fe and Sc layers with thicknesses of $\SI{25}{mm}$ and $\SI{4}{mm}$, respectively.    
\end{itemize} 

The physics data is collected using the trigger signal $\mathcal{S}_{trig}$ from the beam-defining trigger scintillators from point (I), with two additional ECAL selections: a minimum energy deposit of $\mathcal{S}_{\mathrm{PRS}}:=\left(E_{\mathrm{PRS}}\gtrsim\SI{300}{MeV}\right)$ in the PRS and a maximum energy deposit of $\mathcal{S}_{\mathrm{EC}}:=\left(E_{\mathrm{EC}}\lesssim\SI{80}{GeV}\right)$ in the remaining layers of the ECAL. The selection is performed by applying the discrimination on the analog sum of the amplitudes from the $3\times4$ matrix of cells around the central cell. In the following, we refer to this as the production trigger and express it as $\mathcal{S}_{phys}:= \mathcal{S}_{trig}\times\mathcal{S}_{\mathrm{PRS}}\times\mathcal{S}_{\mathrm{EC}}$.

With this sub-detector array, the NA64 experiment can record each particle passing the $\Sphys$ trigger at rates up to $\mathcal{O}(\SI{5}{kHz})$. The magnetic spectrometer selects and measures its momentum, while beam electrons are tagged based on the emitted synchrotron radiation. Upon reaching the target, the ECAL cells record the full pulse waveform, providing a measurement of the e-m shower. The hadronic calorimeters placed after the target allow detection of interactions that could result in secondaries carrying away a significant fraction of the initial beam energy. 

\begin{figure}[ht!]
    \centering
    \includegraphics[width=.99\columnwidth]{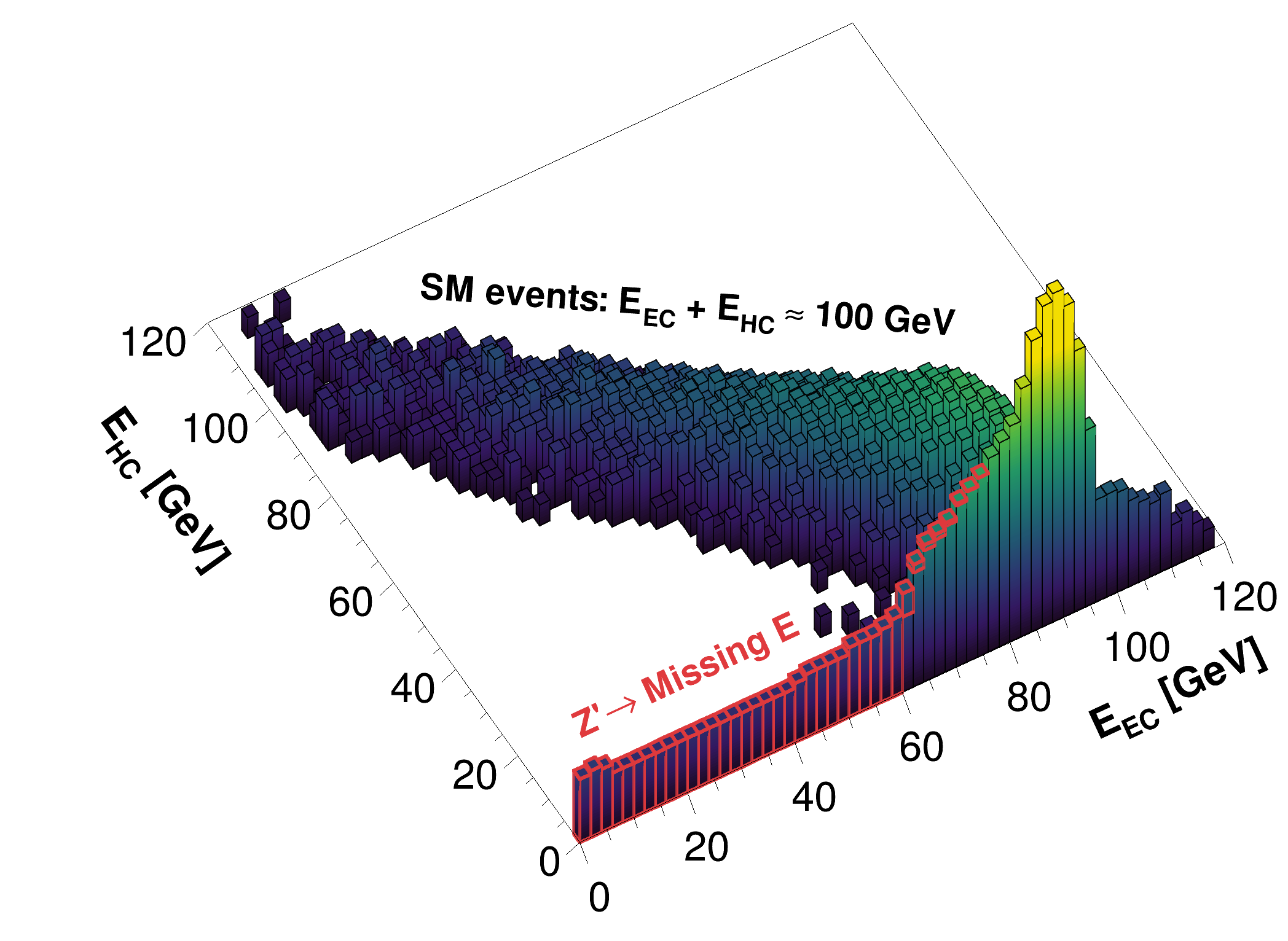}
    \caption{Example of the distribution of physics data events in the $\left(E_{\mathrm{EC}}; E_{\mathrm{HC}}\right)$ plane, including simulated events with production of invisibly-decaying $\Zpr$ in the target with $\mZ=\SI{16.3}{MeV}$ (highlighted in red). Events in the diagonal correspond to SM processes.}
    \label{fig:na64-subdetectors-herm_lego}
\end{figure}

A typical energy distribution in the bi-dimensional ``ECAL vs HCAL" space in NA64, as obtained from a Monte Carlo simulation, is reported in Fig.~\ref{fig:na64-subdetectors-herm_lego}. Here, the highlighted red region illustrates the expected calorimeter response in the case of radiative $\Zpr$ production, which results from the simulation of $\SI{100}{GeV}$ electrons impacting on the ECAL using the Geant4-based Monte Carlo simulation package DMG4 to generate the new vector boson \cite{Agostinelli:2002hh, Allison:2006ve, ALLISON2016186, Oberhauser:2024ozf}.
Such events would leave significant missing energy in the active target, $E_{\mathrm{EC}}$, and no trace of their passage through the HCAL, $E_{\mathrm{HC}}$. In contrast, for most SM interactions, the hadronic calorimeters would detect particles leaking from the target, thus satisfying energy conservation $E_{\mathrm{EC}}+E_{\mathrm{HC}}\approx\SI{100}{GeV}$.

Beyond the events shown in Fig.~\ref{fig:na64-subdetectors-herm_lego}, certain SM processes can lead to background sources with significant missing energy in the ECAL. Hadronic contaminants, such as $\pi^{-}$, $\bar{p}$, and $K^{-}$, may deposit only a fraction of their energy in the ECAL. Additionally, Bremsstrahlung photons produced in the target may interact electromagnetically with the target nuclei, generating a $\mu^{-}\mu^{+}$ pair. These typically penetrate the HCAL modules, depositing energy consistent with two minimum-ionizing particles (MIPs). Lastly, electrons undergoing deep inelastic scattering with the target or upstream detector components may also produce large-angle secondary hadrons that escape detection. For this reason, the experiment relies on the hermetic coverage provided by the hadronic calorimeter (HCAL) modules downstream of the target. A complete description of the expected background is provided in the publication on searches for an invisible dark photon \cite{Andreev:2023uwc}.

\section{Results}\label{sec:Results}

\begin{figure}[!ht]
\centering
\includegraphics[width=0.99\columnwidth]{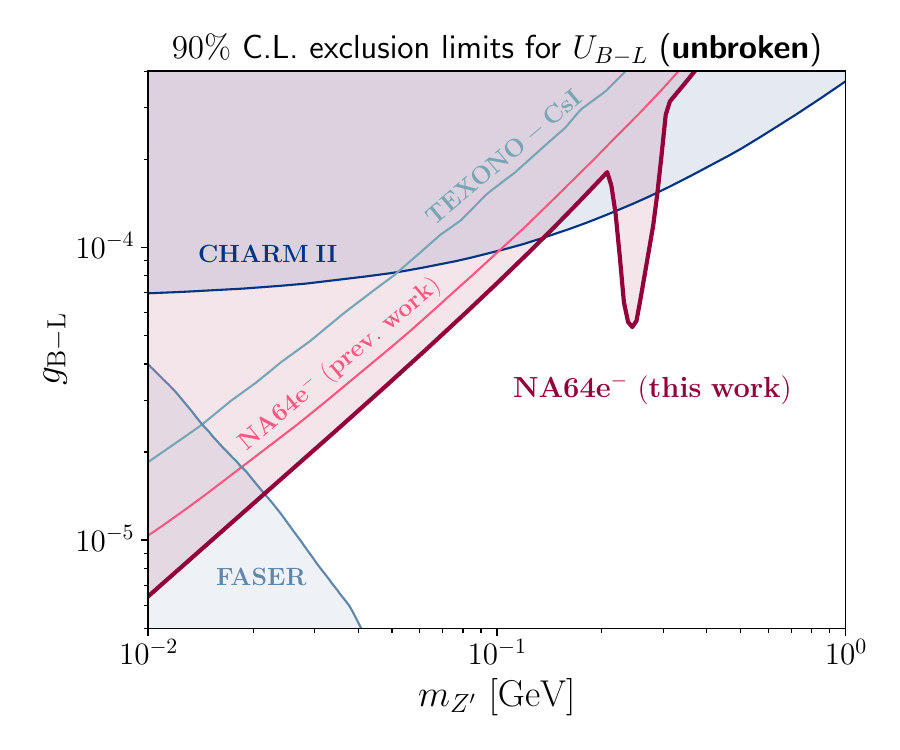}
\includegraphics[width=0.99\columnwidth]{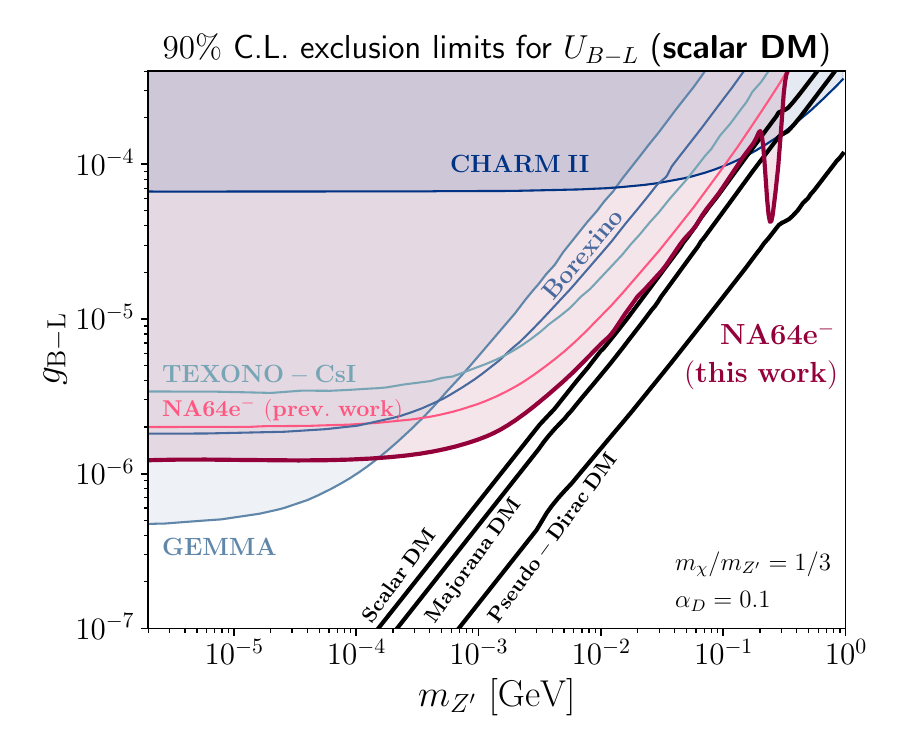}
\caption{$90\%$ C.L. exclusion limits in the $(\mZ, \gBL)$ plane obtained by NA64 for an invisibly decaying $\UBL$ $\Zpr$ using the combined 2016--2022 dataset in the case of an unbroken $B-L$ symmetry with Dirac neutrinos (top) and a direct coupling with a scalar DM particle (bottom). For comparison, the previously published exclusion by NA64 is included \cite{NA64:2022yly}. The DM relic targets have been taken from \cite{Berlin:2018bsc}. }\label{fig:analysis-results-BminusL-limits}
\end{figure}

For this analysis, the combined 2016--2022 dataset, previously analyzed and unblinded in the context of the invisibly-decaying dark photon search \cite{Andreev:2023uwc}, is reinterpreted in terms of the expected missing-energy signature of the $\UBL$ scenario, exploiting similarities between the dark photon and $\Zpr$ phenomenologies. Signal yields are evaluated using DMG4 for both the unbroken symmetry with three light RHN and the DM-coupling case. As no signal was observed in the full statistics of $(9.4\pm0.5)\times10^{11}$ electrons on target, the corresponding $90\%$-C.L. exclusion limits are derived and presented in Fig.~\ref{fig:analysis-results-BminusL-limits}. 

\begin{figure}[!ht]
\centering
\includegraphics[width=0.92\columnwidth]{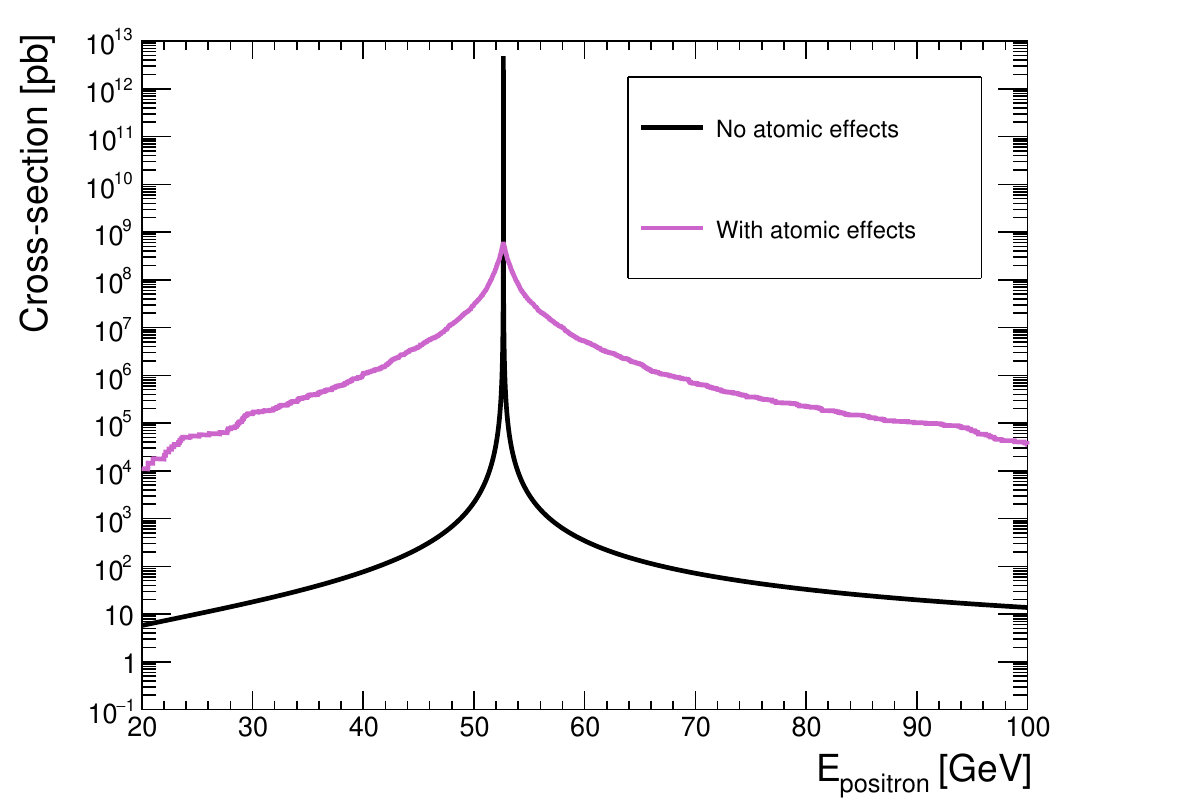}
    \caption{Total cross-section for the resonant annihilation process $e^{+}+e^{-}\rightarrow \Zpr$ as a function of the positron energy $E_{+}$ for $\mZ=\SI{232}{MeV}$ and $\gZ = \SI{3e-3}{}$ in the unbroken case. For these parameters, the decay width is approximately $\Gamma_{\Zpr}\approx\SI{2.5e-1}{keV}$. Displayed are the cases for the cross-section, with and without accounting for the motion of atomic electrons, as implemented in the DMG4 package \cite{Oberhauser:2024ozf}.}\label{fig:analysis-dmg4-atomic-effects}
\end{figure}

Compared with previous NA64 results on $B-L$ models \cite{NA64:2022yly}, this analysis benefits from a threefold increase in statistics through the inclusion of the 2022 dataset and incorporates the resonant-annihilation production channel into the signal simulation. The implementation of this channel follows the strategy adopted for the $L_{\mu} - L_{\tau}$ case, accounting for the effects on the resonant production induced by the motion of atomic electrons in the DMG4 package \cite{Oberhauser:2024ozf}, which leads to a broadening of the otherwise narrow resonance \cite{Andreev:2024lps} as shown in Fig.~\ref{fig:analysis-dmg4-atomic-effects}. This inclusion enhances NA64's sensitivity in the mass range $\mZ\in[200,300]~\mathrm{MeV}$. Note that due to the visible decay channels of $\Zpr$, the exclusion limits set by experiments sensitive to this signature, such as FASER \cite{FASER:2023tle}, also apply in the unbroken case with no DM coupling.

In contrast, in the case of a direct coupling with scalar DM, the decay width is entirely dominated by the channel $\Zpr\rightarrow \chi\bar{\chi}$ for the benchmark case of $\alphaD = Q^{\prime 2}_{B-L, \chi}\gBL^2/4\pi  = 0.1$. Since the $\Zpr$ predominantly decays invisibly, the exclusion limits can be directly derived from the NA64 analysis on the Dark Photon model, as presented in \cite{Andreev:2023uwc}. The limits obtained in the former case need to be rescaled by a factor $e=\sqrt{4\pi\alpha}$ to account for differences in the couplings of the two models. 

\begin{figure}[!ht]
\centering
\includegraphics[width=0.99\columnwidth]{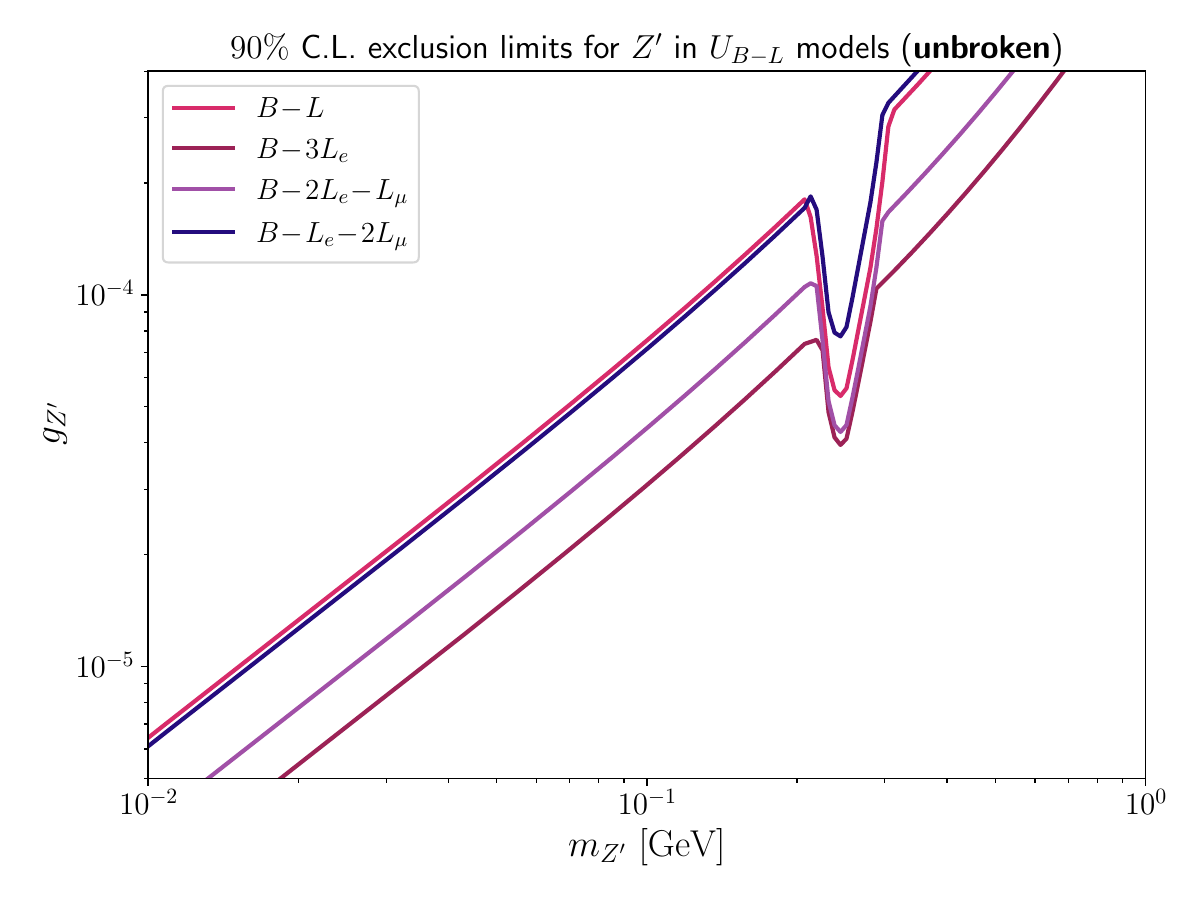}
\caption{$90\%$ C.L. exclusion limits in the $(\mZ, \gZ)$ plane obtained by NA64 for an invisibly decaying $\UBL$ $\Zpr$ using the combined 2016--2022 dataset in all scenarios described in Table~\ref{tab:theory-BminusL-U1prime_models}. }\label{fig:analysis-results-BminusL-limits-four-models}
\end{figure}

To directly compare the NA64 sensitivity across the four different scenarios presented in Table~\ref{tab:theory-BminusL-U1prime_models}, the exclusion contours for all four models are superimposed in Fig.~\ref{fig:analysis-results-BminusL-limits-four-models}. The relative shifts in the exclusion bounds are determined by the specific scaling behavior of the invisible signal from the bremsstrahlung and the resonant annihilation production channel, as mentioned in Section~\ref{sec:Introduction}. The detailed derivation of the resonant annihilation scaling is provided in the Appendix~\ref{sec:appendix-Zpr-resonant}.

\section{Conclusions}\label{sec:Conclusions}

We have presented new constraints on a light vector mediator associated with a gauged $\UBL$ symmetry using the full electron-beam dataset collected by the NA64 experiment between 2016 and 2022. The results probe the sub-GeV mass range and extend previous laboratory bounds on the coupling strength $g_{B-L}$.

A key improvement of this analysis is the inclusion of the resonant $e^{+}e^{-}$ annihilation production channel, which enhances the sensitivity of NA64 in the mediator mass range $\mZ \in [200,300]~\mathrm{MeV}$. Together with the larger integrated statistics from the full 2016--2022 dataset, this leads to improvements over earlier NA64 results. In the unbroken $\UBL$ scenario, the resulting limits surpass those derived from neutrino--electron scattering experiments and constitute the most stringent laboratory constraints on $g_{B-L}$ below the GeV scale.

The limits can be straightforwardly reinterpreted for a broad class of anomaly-free $B-n_{i}L_{i}$ models by rescaling the fermion charges entering the production cross sections and decay widths. In scenarios where the $\Zpr$ couples directly to dark matter, its decay width is dominated by invisible channels, and the corresponding exclusion limits can be directly obtained from the NA64 invisible-mode analysis.

These results demonstrate the strong sensitivity of missing-energy fixed-target experiments to light gauge bosons with direct couplings to Standard Model fermions and their complementarity with neutrino scattering experiments. The inclusion of resonant production mechanisms highlights the synergy between different production modes and further strengthens NA64's role in exploring well-motivated extensions of the Standard Model that are connected to neutrino mass generation and dark sectors. This program is complemented by NA64 running in muon mode, which probes a broader range of models beyond the electron-coupling sector \cite{NA64:2024nwj}.

\section*{Acknowledgments}
We gratefully acknowledge the support of the CERN management and staff, as well as the technical staff of the participating institutions, for their vital contributions. This work was supported by the HISKP, University of Bonn (Germany), ETH Zurich Grant No. 22-2 ETH-031, and SNSF Grant No. 186181, No. 186158, No. 197346, No. 216602, No. 232705 (Switzerland), and FONDECYT (Chile) under Grant No. 1240066 and Grant No. 3230806, and ANID - Millenium Science Initiative Program - ICN2019 044 (Chile), and RyC-030551-I and PID2021-123955NA-100 funded by MCIN/AEI/ 10.13039/501100011033/FEDER, UE (Spain), and COST Action COSMIC WISPers CA21106, supported by COST (European Cooperation in Science and Technology). This result is part of a project funded by the European Research Council (ERC) under the European Union's Horizon 2020 research and innovation programme, Grant agreement No. 947715 (POKER). This work is partially supported by ICSC -- Centro Nazionale di Ricerca in High Performance Computing, Big Data and Quantum Computing, funded by the European Union -- NextGenerationEU.

\appendix

\section{Calculations for the resonant production of \texorpdfstring{\ensuremath{\Zpr}}{Zpr}}\label{sec:appendix-Zpr-resonant}

A new $\Zpr$ vector boson can be produced in NA64 by direct annihilation of an electron-positron pair, and the cross-section of this process is peaked when the invariant mass $\sqrt{s}$ of the pair is the same as the $\Zpr$ mass. As a result of the massive nature of $\Zpr$, the process contains a Breit-Wigner term that describes the resonant behavior of the cross-section. Similar to the case of the SM Z boson, the cross-section for the annihilation into any valid fermion-antifermion pair can be written as:
\begin{equation}
    \sigma(e^{+}e{-}\rightarrow\Zpr\rightarrow f \bar{f}) = \frac{12\pi s}{\mZ^{2}}\frac{\Gamma_{e^{+}e^{-}}\Gamma_{f\bar{f}}}{(s-\mZ^{2})^{2}+\mZ^{2}\Gamma^{2}_{\mathrm{tot}}}
    \label{eq:appendix-Zpr-resonant-XS}
\end{equation}

where $s = 2m_{e}E_{e^{+}}$ if we consider the atomic electrons at rest, and $\Gamma_{ff}$ describes the decay width for the different final states of the SM fermions coupling to $\Zpr$. The term $E_{e^{+}}$ corresponds to the energy of secondary positrons from the e-m shower. In particular, $\Gamma_{\mathrm{tot}}$ denotes the total decay width of the $\Zpr$. In the scenario considered in this work, the resonance widths are extremely narrow, meaning the motion of the atomic electrons can no longer be neglected. To accurately model the complete cross-section, including this kinematic broadening, we utilize the recent developments implemented in the Geant4-compatible Dark Matter event generator, DMG4 \cite{Oberhauser:2024ozf}.
The cross section reaches its maximum at $\sqrt{s}=m_{\Zpr}$, yielding
\begin{equation}
\sigma_{\mathrm{max}}(e^{+}e^{-}\rightarrow\Zpr\rightarrow f \bar{f}) =
12\pi
\frac{\Gamma_{e^{+}e^{-}}\Gamma_{f\bar{f}}}{\mZ^{2}\Gamma^{2}_{\mathrm{tot}}}.
\label{eq:appendix-Zpr-resonant-XS-max}
\end{equation}

If we consider the total annihilation cross section summed over all allowed SM fermion
pairs $f\bar f$, the partial widths in the numerator sum to the total width,
\begin{equation}
\sum_f \Gamma_{f\bar f} = \Gamma_{\mathrm{tot}}
\end{equation}
The maximum cross-section, therefore, scales as
\begin{equation}
\sigma^{\mathrm{tot}}_{\mathrm{max}} \propto
\frac{\Gamma_{e^{+}e^{-}}}{\Gamma_{\mathrm{tot}}}
= \mathrm{Br}(\Zpr \rightarrow e^{+}e^{-})
\end{equation}

Thus, the peak cross-section depends only on the branching ratio of the mediator to
electrons, which is illustrated in Fig.~\ref{fig:theory-BminusL-Zpr_decay_BR}. Additionally, $\sigma^{\mathrm{tot}}_{\mathrm{max}}$ is independent of the overall coupling strength, since
$\Gamma_{e^{+}e^{-}} \propto \gZ^{2}$ and $\Gamma_{\mathrm{tot}} \propto \gZ^{2}$.

The partial decay widths to the different combinations of fermions are clearly dependent on the coupling choices in each particular  $B-n_{i}L_{i}$-type model. For the most general case, we can write for fermions:
\begin{equation}
    \Gamma_{f\bar{f}} = \frac{1}{3}\mZ\frac{Q^{\prime 2}_{f}\gZ^{2}}{4\pi}\left(1+2\frac{m_{f}^2}{\mZ^{2}}\right)\sqrt{1 - 4\frac{m^2_{f}}{\mZ^2}}
    \label{eq:appendix-Zpr-resonant-decay-width-ff}
\end{equation}

Analogously, the decay width for a scalar DM particle pair, denoted as $\chi_{s}$, can be written including the appropriate phase space suppression term and an additional factor $\frac{1}{4}$, which results from the lower number of degrees of freedom compared to a spin $\frac{1}{2}$ particle:
\begin{equation}
    \Gamma_{\chi_{s}\bar{\chi_{s}}} = \frac{1}{12}\mZ\frac{Q^{\prime 2}_{\chi_{s}}\gZ^{2}}{4\pi}\left(1-4\frac{m_{\chi_{s}}^2}{\mZ^{2}}\right)^{3/2}
    \label{eq:appendix-Zpr-resonant-decay-width-ss}
\end{equation}

Let us now consider the NA64 case, where the $\Zpr$ mass is constrained to the interval $\mZ\in[200,300]~\mathrm{MeV}$. These sensitivity bounds are given by the missing-energy threshold in the ECAL $E^{miss}_{\mathrm{EC}}$ and the primary beam particle energy. In this case, $m_{e} \ll \mZ$, so that the electron channel is essentially 
unsuppressed, while the phase-space suppression factor remains relevant only for the muon channel \footnote{If the coupling to DM is considered, then the partial width in Eq.~\ref{eq:appendix-Zpr-resonant-decay-width-ss} must also be included in the total width.}. We can therefore see that for this approximation and substituting in the expressions for the partial widths, the peak of the total cross-section can be written as:

\begin{equation}
  \sigma_{\rm max}^{\rm tot} \simeq
  \frac{12 \pi}{m^2_{Z'}} \,
  \Big(R_{\nu\bar\nu} + R_{e^-e^+} + R_{\mu^-\mu^+}\Big)^{-1} 
  \,,
  \label{eq:appendix-Zpr-resonant-XS-max-tot}
\end{equation} 

where $R_{\ell\bar\ell}$ are the partial contributions of
the $Z'$ annihilation into leptonic pairs:
$R_{\nu\bar\nu} = 1 + (Q^{\prime}_\mu/Q^{\prime}_e)^2 + (Q^{\prime}_\tau/Q^{\prime}_e)^2$ is
the combined effect of three neutrino-antineutrino channels,
$R_{e^-e^+} = 1$ and
\begin{equation}
  R_{\mu^-\mu^+} =
  \Big(\frac{Q^{\prime}_\mu}{Q^{\prime}_e}\Big)^2 \,
  \Big(1 + 2 \frac{m_\mu^2}{m^2_{Z'}}\Big) \,
  \sqrt{1 - 4 \frac{m_\mu^2}{m^2_{Z'}}}
  \label{eq:appendix-Zpr-resonant-XS-width-mumu}
\end{equation}\smallskip

are the $e^-e^+$ and $\mu^-\mu^+$ contributions, respectively. 

An important observation is that the resonant annihilation cross-section depends only on the ratios of the charges under the new interaction $Q^{\prime}_{i}$. Therefore, and given that the exclusion limits in the $(\mZ, \gZ)$ parameter space scale as $\gZ \propto 1/\sqrt{n_{s}} \propto 1/\sqrt{\sigma}$, the resulting signal yields from the annihilation channel in the $B-L$ model can be reinterpreted to any variation of a $B-n_{i}L_{i}$ symmetry by rescaling the corresponding partial and total widths. This behaviour contrasts with the radiative $\Zpr$ production in NA64, which scales directly with the electron coupling as $\sigma_{Brem} \propto Q^{\prime 2}_{e}\gZ^{2}$.

Finally, for $\alphaD = 0.1$, the corresponding charge of the DM particles is comparatively large and effectively means that the DM decay almost exclusively determines the cross-section and branching ratios. Therefore, the sensitivity of NA64 to this particular model can be directly obtained from the invisible $\Apr$ case, by substituting $e\varepsilon\leftrightarrow \gZ$.

\bibliography{bibliographyNA64_inspiresFormat,bibliographyNA64exp_inspiresFormat,bibliographyOther_inspiresFormat}

\end{document}